# Use of Groundwater Lifetime Expectancy for the Performance Assessment of a Deep Geologic Waste Repository: 1. Theory, Illustrations, and Implications

F. J. Cornaton[1, 3], Y.-J. Park[1*], S. D. Normani[2],

E. A. Sudicky[1] and J. F. Sykes[2]

1: Department of Earth and Environmental Sciences
University of Waterloo
Waterloo, Ontario, Canada N2L 3G1
E-mail: yj2park@sciborg.uwaterloo.ca,
sudicky@sciborg.uwaterloo.ca,
Fax: 1-519-746-7484

2: Department of Civil and Environmental Engineering
University of Waterloo
Waterloo, Ontario, Canada N2L 3G1
E-mail: sdnorman@uwaterloo.ca,
sykesj@uwaterloo.ca

3: Now at:
CHYN, Centre of Hydrogeology, University of Neuchâtel
Emile-Argand 11, CP 158
CH-2009 Neuchâtel, Switzerland
fabien.cornaton@unine.ch

[*] Corresponding author

**REVISED VERSION**

# Abstract

Long-term solutions for the disposal of toxic wastes usually involve isolation of the wastes in a deep subsurface geologic environment. In the case of spent nuclear fuel, if radionuclide leakage occurs from the engineered barrier, the geological medium represents the ultimate barrier that is relied upon to ensure safety. Consequently, an evaluation of radionuclide travel times from a repository to the biosphere is critically important in a performance assessment analysis. In this study, we develop a travel time framework based on the concept of groundwater lifetime expectancy as a safety indicator. Lifetime expectancy characterizes the time that radionuclides will spend in the subsurface after their release from the repository and prior to discharging into the biosphere. The probability density function of lifetime expectancy is computed throughout the host rock by solving the backward-in-time solute transport adjoint equation subject to a properly posed set of boundary conditions. It can then be used to define optimal repository locations. The risk associated with selected sites can be evaluated by simulating an appropriate contaminant release history. The utility of the method is illustrated by means of analytical and numerical examples, which focus on the effect of fracture networks on the uncertainty of evaluated lifetime expectancy.

# 1. Introduction

Waste disposal practices have evolved substantially over the last twenty years. Engineers and scientists play an essential role by providing scientific insight into the long-term performance and safety of repositories for spent nuclear fuel. Currently, the preferred option for the long-term storage of radioactive waste is to isolate it in a deep geologic environment. The potential suitability of a candidate site to host a geologic repository will rely, in part, on a demonstrated understanding of the groundwater flow system and its evolution over time scales of perhaps tens of thousands of years or longer. The characterization of a flow system is achieved through iterative site characterization processes in which multi-disciplinary lithostructural, hydrogeologic, paleohydrogeologic, geophysical, hydrogeochemical, and geomechanical data are gathered at increasing levels of detail to support development of a conceptual flow system model. Numerical modeling offers a systematic framework in which these site specific data may be integrated and then assessed for consistency, to instill more confidence in the conceptual flow system model(s) and provide a basis for predictive modeling of long-term flow system evolution and repository performance.

The safety of the subsurface waste repositories depends on two main barriers: the engineered barrier and the natural geological barrier. In the case that leakage occurs from the engineered barrier, the geological medium represents the ultimate barrier that can ensure safety. Consequently, the release of toxic contaminants from the repository and their migration within the groundwater flow system to the biosphere must, therefore, be considered. The two main aspects associated with the analysis of the safety of subsurface waste repositories can be formalized by the following: the repositories must ensure maximum travel times from the waste site to the biosphere as well as a maximum degree of dilution of the released contaminants [*Davies et al.*, 1991; *Birdsell et al.*, 2000; *Pohlmann et al.*, 2006]. Moreover, particular emphasis needs to be given to reservoir structure and parameter distribution uncertainty and its impact on radionuclide transport [*Davies et al.*, 1991; *Selroos et al.*, 2002].

In this study we develop a framework for the analysis of the safety of deep geologic repositories based on specific mathematical modeling approaches. As the main safety indicator, we introduce the concept of lifetime expectancy, which is defined as the time required by water molecules to reach an outlet of a groundwater system [*Cornaton and Perrochet*, 2006a and b]. In an appropriate representative elementary volume, the variable lifetime expectancy is a random variable characterized by a probability density function. Here we focus on radionuclide lifetime expectancy by evaluating its travel time after release in groundwater at the repository locations, and prior to reaching the biosphere. Travel time probabilities have been the subject of great interest in many studies characterizing solute transport in subsurface hydrology [*Jury*, 1982; *Dagan*, 1982, 1987; *Shapiro and Cvetkovic*, 1988; *Dagan*, 1989; *Dagan and Nguyen*, 1989; *Jury and Roth*, 1990; *Goode*, 1996; *Varni and Carrera*, 1998; *Ginn*, 1999], as well as the migration of radionuclides [*Andričevíc et al.*, 1994]. The travel time probability is commonly defined as the response function to an instantaneous unit flux impulse [*Danckwerts*, 1953; *Jury*, 1982].

In the present work, the radionuclide lifetime expectancy is computed by solving the backward-in-time solute transport adjoint equation with a set of properly posed boundary conditions. The backward-in-time approach can provide the information on lifetime expectancy from any arbitrary location to the biosphere, by reversing the time and subsurface velocities and simulating radionuclide transport from the biosphere towards the subsurface. The radionuclide lifetime expectancy is analyzed to define potential and optimal repository locations. The risk associated with selected sites can then be evaluated by incorporating an appropriate contaminant release history. In order to illustrate the utility of the methodology, we derive analytical solutions for temporal moments of lifetime expectancy in a system of parallel vertical fractures embedded in a porous matrix, and by means of theoretical numerical examples, investigate the effect of fracture network geometry on lifetime expectancy as a safety indicator for potential repository sites.

## 2. Theory

There are many social, economic, and scientific factors to be considered for radioactive waste disposal [*US EPA*, 1999; *Vieno*, 1999; *Walker et al.*, 2001; *NRC*, 2005]. Amongst classical scientific criteria used to determine the optimal location of an underground waste repository are: (1) the longest travel time from the waste site to the biosphere, (2) the least dose (or maximum dilution) to the biosphere, and (3) the minimal prediction uncertainty. In the following sections, we mainly focus on points (1) and (2). Uncertainties related to subsurface hydrogeological parameter distributions will be discussed in a separate article in the framework of the Canadian Shield environment as a potential repository host formation [*Park et al.*, 2007]. We first recall the equivalence between the standard advection-dispersion equation (ADE) and standard diffusion theory that relates the dynamics of a diffusion process to the Fokker-Planck equation. The lifetime expectancy of water molecules is then defined by introducing the formal adjoint of the forward equation.

## 2.1. Forward and Backward-in-Time Transport Equations

The stochastic motion of particles in dynamic systems has been studied extensively in the theory of stochastic differential equations. The spreading of a contaminant mass can be described by the random motion of solute particles, and the ADE is assimilated to the forward Fokker-Planck (or forward Kolmogorov) equation. Therefore, it is possible to derive an Îto stochastic differential equation as a random walk model for the movement of a contaminant particle that is exactly consistent with the advection-dispersion model. Two alternative differential equations have been developed (see *e.g. Gardiner*, 1983): the forward Fokker-Planck equation (FPE) and the backward-in-time Kolmogorov equation (BKE) [*Kolmogorov*, 1931]. The use of forward and backward models for location and travel time probability has become a classical mathematical approach for contaminant transport characterization and prediction [*Uffink*, 1989; *LaBolle et al.*, 1998; *Varni and Carrera*, 1998; *Neupauer and Wilson*, 1999; *Spivakovskaya et al.*, 2005]. The forward version of the Fokker-Planck equations points to the future, and the transformation from probabilities and particle densities to a physical concept of a solute concentration is straightforward. In the backward formulation, the correspondence between probability and concentration is lost. When a particle is observed at a certain time and position, its possible location at an earlier time covers an area that increases in size as we look further back in time [*Uffink*, 1989].

A Lagrangian approach for solute transport describes the spreading of a contaminant in a flow system. The position of a particle $X_{x_0,t_0}(t)$, that was released at time $t_0$ at $X_{x_0,t_0}(t_0) = x_0$ is described by the following Îto stochastic differential equation (SDE):

$$dX_i = a_i(t,X)dt + \sigma_{ij}(t,X)dW_j(t) \qquad (1)$$

where $a_i$ is a vector function, $\sigma_{ij}$ is a matrix function related to the dispersion tensor, and $W_j$ is a vector-valued Brownian motion process [*Gardiner*, 1983]. The transition density $p(x,t|x_0,t_0)$ to find a particle at position $x$ at time $t$, given that it was released at time $t_0$ at position $x_0$, can be obtained by solving the Fokker-Planck equation (FPE) [*Langevin*, 1908].

$$\frac{\partial p}{\partial t} = -\frac{\partial a_i p}{\partial x_i} + \frac{\partial^2 b_{ij} p}{\partial x_i \partial x_j} \qquad (2)$$

where the terms $a_i = a_i(x,t)$ and $b_{ij} = b_{ij}(x,t) = \sigma_{ik}\sigma_{jk}$ denote the drift vector and noise tensor, respectively. By defining $a_i = v_i + \partial b_{ij}/\partial x_j$ and $b_{ij} = D_{ij}$ in terms of the pore velocity $v_i$ and the dispersion tensor $D_{ij}$ and by replacing the variable $p(x,t)$ with $\phi(x)C(x,t)$, where $\phi$ is porosity, or mobile water content and $C$ is the solute resident concentration, one can show that (2) becomes equivalent to the following classical forward advection-dispersion equation (ADE):

$$\frac{\partial \phi C}{\partial t} = -\frac{\partial}{\partial x_i}(q_i C) + \frac{\partial}{\partial x_i}\phi D_{ij}\frac{\partial C}{\partial x_j} \qquad (3)$$

where $q_i = \phi v_i$ is the water flux vector and the dispersion coefficient tensor $D_{ij}$ is defined by *Bear* [1972].

Assuming that the flow field is divergence-free and at steady state, the BKE can be expressed by the following:

$$\frac{\partial p}{\partial \tau} = -\frac{\partial a_i^* p}{\partial x_i} + \frac{\partial^2 b_{ij} p}{\partial x_i \partial x_j} \qquad (4)$$

where $\tau = t_0 - t$ is backward (or reverse) time, and the drift vector $a_i^*$ and noise tensor $b_{ij}$ are defined as follows:

$$a_i^*(\boldsymbol{x}) = -v_i(\boldsymbol{x}) + \phi(\boldsymbol{x})^{-1}\frac{\partial}{\partial x_j}\left(\phi(\boldsymbol{x})b_{ij}(\boldsymbol{x})\right) \qquad (5a)$$

$$b_{ij}(\boldsymbol{x}) = D_{ij}(\boldsymbol{x}) \qquad (5b)$$

Equation (4) has the form of a forward equation, showing that the FPE and the BKE are equivalent to each other. Only the drift coefficient differs from its original definition: velocity presents a reversed sign to handle the backward-in-time evolution. Equation (4) is often called the formal adjoint of the FPE [*Garabedian*, 1964; *Arnold*, 1974]. Suppose that a particle is found at the position $\boldsymbol{x}_0$ at time $t_0$. To evaluate the probability that this particle was at an upstream position at an earlier time, one can solve the BKE as expressed by (4), with the initial condition $p(\boldsymbol{x},t_0|\boldsymbol{x}_0,t_0) = \delta(\boldsymbol{x}-\boldsymbol{x}_0)$. Solutions of the FPE, subject to such an initial condition and any appropriate boundary conditions, yield solutions of the BKE as well. For the FPE, solutions exist for $t \geq t_0$ with ($x_0, t_0$) fixed, while, for the BKE, solutions exist for $t \leq t_0$, so that the backward equation expresses the development in $t_0$, and $p(x, t_0 | x_0, t_0) = \delta(x - x_0)$ is rather termed as the final condition [*Gardiner*, 1983]. The forward equation gives more directly the values of measurable quantities (such as concentration $C$) as a function of the observed time $t$. The backward equation finds most applications in the study of first passage time or exit problems, in which we find the probability that a particle leaves a region in a given time. Using the definitions (5), and replacing $p = \phi g$ in (4), one can obtain:

$$\frac{\partial \phi g}{\partial \tau} = \frac{\partial}{\partial x_i}(q_i g) + \frac{\partial}{\partial x_i} \phi D_{ij} \frac{\partial g}{\partial x_j} \tag{6}$$

where $g = g(x, t)$ is a backward-in-time probability density for the particle location.

Equation (6) corresponds to the so-called "backward-in-time" ADE [*Uffink*, 1989; *Van Kooten*, 1995; *Wilson and Liu*, 1997; *Neupauer and Wilson*, 1999; *Weissmann et al.*, 2002; *Cornaton and Perrochet*, 2006a, b]. The probability density function (PDF) for lifetime expectancy can be obtained by solving (6) when a unit pulse flux input is uniformly applied over the surface discharge areas of the ground water reservoir. Specific details concerning the numerical implementation of the lifetime expectancy boundary value problem can be found in *Cornaton and Perrochet* [2006a]. In the current problem of radioactive waste repository safety analysis, one can consider that the target to be protected is known, and this target can be taken as the union of all the outlets for the region of concern, as the outflow of radioactivity at these outlets can potentially harm the biosphere.

The repository radioactive source is *a priori* unknown since we are investigating potentially acceptable repository locations.

**2.2. Mean Lifetime Expectancy as a Safety Indicator**

Lifetime expectancy is given as a PDF for an appropriate representative volume and the mean lifetime expectancy, first temporal moment, is an average expected lifetime for the volume as a representative statistic for the PDF. Since this statistic can be more comparable to radiometric data and be computed with relative ease, it may be used for preliminary analysis in selecting the safest possible repository locations. Note, however, that a longer mean lifetime expectancy does not always guarantee the least dose, or the later arrival of risks to the biosphere, as the averaged travel time of contaminants can be much longer than first or peak arrivals in multiple pathways reactive systems, even though it might so indicate. Similar formulations and arguments can be found in *Goode* [1996] and *Ginn* [1999] where they derive transport equation of groundwater mean age within forward-in-time framework.

Suppose that $g(x,\tau)$ is a solution of equation (6), given that $g(x,0) = \dfrac{\delta(x - x_b)}{\phi}$, for all $x_b$ on the outlet boundary. The density function $g(x,\tau)$ is equivalent to the lifetime expectancy of a water molecule at the location $x$, prior to exiting the reservoir. The mean lifetime expectancy $E(x)$ is defined as the first temporal moment of the function $g(x,\tau)$:

$$E(x) = \langle g(x,\tau) \rangle = \int_0^\infty \tau g(x,\tau) d\tau \qquad (7)$$

Multiplying (6) by backward time $\tau$ and integrating through all time gives

$$\int_0^\infty \left(\tau \frac{\partial \phi g}{\partial \tau}\right) d\tau = \frac{\partial}{\partial x_i} \cdot q_i \left(\int_0^\infty \tau g d\tau\right) + \frac{\partial}{\partial x_i} \cdot \phi D_{ij} \frac{\partial}{\partial x_j}\left(\int_0^\infty \tau g d\tau\right) \tag{8}$$

By integrating by parts the left side of (8), the function $E(x)$ is the solution of the first moment form of (6), given that $E(x_b) = 0$:

$$\frac{\partial}{\partial x_i}(q_i E) + \frac{\partial}{\partial x_i} \phi D_{ij} \frac{\partial E}{\partial x_j} + \phi = 0 \tag{9}$$

where $\int_0^\infty g d\tau = 1$ and $g(\mathbf{x},0) = g(\mathbf{x},\infty) = 0$. Note that for given groundwater flow system, mean lifetime expectancy is an adjoint state of the mean age – first temporal moment of age PDF as formulated by *Goode* [1996] and *Ginn* [1999]. In (9), mean lifetime expectancy is continuously generated during groundwater flow, since porosity $\phi(x)$ acts as a source term. This source term indicates that groundwater is aging one unit per unit time, on average. The mean lifetime expectancy equation can be easily handled by numerical codes that solve ADEs, by distributing a source term equal to porosity, and by reversing the velocity field.

**2.3. Mass Loads at the Biosphere and Risk Assessment**

In order to simulate the transient history of a contaminant release with a relatively low computational burden, we again make use of the backward-in-time equation. The solution of (6), $g(x,t)$, is the PDF for the lifetime expectancy of water molecules. It can be shown that the total net mass flux $j_b$ at the biosphere, resulting from a transient input of intensity $m^*(\boldsymbol{x},t)$ [MT$^{-1}$] at the repository site location, can be evaluated by the following equation [*Cornaton*, 2003]:

$$j_b(t) = \int_\Omega \left( \int_0^t m^*(\boldsymbol{x}, t-u) g(\boldsymbol{x}, u) du \right) \delta(\boldsymbol{x} - \boldsymbol{x}_r) d\boldsymbol{x} \tag{10}$$

where $\boldsymbol{x}_r$ indicates the coordinates over the repository locations. Equation (10) is similar to the one presented in *Rubin* [2003], but it presents the considerable advantage that, since $g(\boldsymbol{x},t)$ needs to be solved only once, the mass outflux $j_b(t)$ (and thus the associated risk) can be post-processed for a series of different repository locations. The flux-averaged concentration at the biosphere can be obtained by normalizing $j_b(t)$ with the outflow discharge rate. A decay process can easily be added in the formulation (10) by substituting the lifetime expectancy density $g(\boldsymbol{x},t)$ by the defective density $g_d(\boldsymbol{x},t) = \exp(-\frac{\ln 2}{\omega} t) g(\boldsymbol{x},t)$, where $\omega$ is the radionuclide half-life and the defective density represents the probability density that does not conserve the mass and thus the total probability of which is less than one [*Andričevic et al.*, 1994]. Equation (10) can then be applied to a series of specific radionuclides with different half-lives.

## 3. Lifetime Expectancy in Fractured Porous Media

As the subsurface disposal of nuclear fuel wastes aims to store the waste for a sufficiently long time to ensure safety (tens of thousands of years or longer), the deep geologic environment has been considered as a host for such repositories, with the Deep Geologic Repository Technology Program of Ontario Power Generation, Canada being an example. In a crystalline geologic environment, interconnected permeable fracture networks have attracted concern since they could act as pathways for rapid contaminant migration. In this section, we derive analytical solutions for first and second temporal moments of lifetime expectancy in a semi-infinite domain with a set of parallel vertical fractures in a porous matrix block, which could represent a typical fractured crystalline environment (for example, *Sudicky and Frind*, 1982) (Figure 1).

To derive analytical solutions for lifetime expectancy, fractures with a constant aperture $2b$ are distributed with equal spacing $L$ (Figure 1). Fluid flow is assumed to be in an upward direction with a constant flow rate $q_f$ in each fracture. Transport in fractures is assumed to be one-dimensional advective-dispersive along the fracture axis, while diffusion is a dominant transport process in the matrix block perpendicular to the fracture axis. The BKE is derived for the fracture and matrix domains from (6) as follows:

$$\frac{\partial \phi_f g_f}{\partial \tau} = \frac{\partial}{\partial z}(q_f g_f) + \frac{\partial}{\partial z}\phi_f D_f \frac{\partial g_f}{\partial z} - \frac{\phi_m D_m}{b}\frac{\partial g_m}{\partial x}\bigg|_{x=0} \tag{11}$$

$$\frac{\partial \phi_m g_m}{\partial \tau} = \frac{\partial}{\partial x}\phi_m D_m \frac{\partial g_m}{\partial x} \tag{12}$$

where $x$ ($[0,\infty)$) and $z$ ($[0,L]$) are spatial coordinates along and perpendicular to the fracture axis, respectively, $g$ is lifetime expectancy probability density, $\phi$ is porosity, $D$ is effective dispersion coefficient, and subscripts $f$ and $m$ denote fracture and matrix respectively. Symmetry and continuity constrain the boundary conditions as follows:

$$g_f(z=0,\tau>0) = 0 \tag{13a}$$

$$g_m(x=0,z,\tau) = g_m(x=L,z,\tau) = g_f(z,\tau) \tag{13b}$$

$$\left.\frac{\partial g_m}{\partial x}\right|_{x=\frac{L}{2}} = 0 \tag{13c}$$

Note that the continuity of mass flux is applied to (11) explicitly and (13b) implies the continuity of lifetime expectancy PDF between fracture and matrix domains. Similarly to the derivation of (9) from (6), equations for mean life expectancy in fracture and matrix domains can be derived from (11) and (12) as follows:

$$-q_f \frac{\partial E_f}{\partial z} + \phi_f D_f \frac{\partial^2 E_f}{\partial z^2} + \phi_f = -\frac{\phi_m D_m}{b} \left.\frac{\partial E_m}{\partial x}\right|_{x=0} \tag{14}$$

$$\phi_m D_m \frac{\partial^2 E_m}{\partial x^2} + \phi_m = 0 \tag{15}$$

where $E$ is mean lifetime expectancy. The boundary conditions in (13) can also be transformed as follows:

$$E_f(z=0) = 0 \tag{16a}$$

$$E_m(x=0, z) = E_m(x=L, z) = E_f(z) \tag{16b}$$

$$\left.\frac{\partial E_m}{\partial x}\right|_{x=\frac{L}{2}} = 0 \tag{16c}$$

The solution for (14) with (16) is given by

$$E_f(z) = \frac{z}{2bq_f}\left(\phi_m L + 2b\phi_f\right) = \frac{z}{v_f}\left(\frac{\phi_m L}{2b\phi_f} + 1\right) \tag{17}$$

where $v_f = q_f / \phi_f$ is the pore velocity in the fractures. The resulting mean lifetime expectancy in the fractures is a linear function of depth $z$ and it becomes greater with matrix diffusion as the volume of water in the matrix block ($\phi_m L$) is larger than the volume of water in the fractures ($2b\phi_f$). On the other hand, when the matrix porosity is negligible ($\phi_m L \ll 2b\phi_f$), the mean lifetime expectancy at depth $z$ becomes the trivial piston-flow solution $z/v_f$. The mean lifetime expectancy in the matrix block is also given by:

$$E_m(x,z) = -\frac{1}{2D_m}x(x-L) + E_f(z) \tag{18}$$

The solution (18) shows that the mean lifetime expectancy is parabolic across the matrix block constrained by the solution in the bounding fractures $E_f(z)$. Maximum mean lifetime expectancy is obtained in the midpoint of the block ($x = L/2$) as $L^2/8D_m + E_f(z)$, indicating that it becomes greater as the block size becomes larger and with a smaller effective matrix diffusion coefficient (Figure 2).

As solute travels, it spreads wider, and the variance of lifetime expectancy increases as the mean lifetime expectancy increases. The variance represents the uncertainty of the mean as a safety indicator for repositories: for instance, if the mean of the lifetime expectancy is large and its variance is small for a certain location, it could be considered to be a safe zone, while if the variance is large compared to the mean, a long lifetime expectancy may not guarantee its safety. In this context, we derive the variance of lifetime expectancy for a fracture as:

$$\sigma_f^2(z) = V_f(z) - E_f^2(z) = \int_0^\infty (\tau - E_f(z))^2 g_f(z,\tau) d\tau \tag{19}$$

where $V_f(z) = \int_0^\infty \tau^2 g_f(z,\tau) d\tau$. A second moment equation similar to (14) and (15) can be derived by multiplying (13) by $\tau^2$ and by integrating by parts as follows:

$$-q_f \frac{\partial V_f}{\partial z} + \phi_f D_f \frac{\partial^2 V_f}{\partial z^2} + 2\phi_f E_f = -\frac{\phi_m D_m}{b} \left.\frac{\partial V_m}{\partial x}\right|_{y=b} \quad (20)$$

$$\phi_m D_m \frac{\partial^2 V_m}{\partial x^2} + 2\phi_m E_m = 0 \quad (21)$$

The solution for the variance of lifetime expectancy in fractures $\sigma_f^2(z)$ has a form of $\alpha z^2 + \beta z$, where the coefficients $\alpha$ and $\beta$ are functions of $b$, $L$, $\phi$, $D$ and $q_f$. To avoid the redundancy but for illustrative purposes, the solutions $E_f(z)$, $\sigma_f(z)$, and $\sigma_f/E_f$ are plotted with depth for the case when $L$ = 150m, $b$ = 20μm, $\phi_f$ = 0.01, $\phi_m$ = 0.001, $q_f$ = 10 m/year, $D_f$ = $D_m$ = 0.00725 m²/year (Figure 3a). Figure 3a shows that the mean and standard deviation for lifetime expectancy become greater with depth. Note that increasing the mean and variance in lifetime expectancy has both positive and negative effects for hosting a repository as we can expect a longer travel time from the repository location, but with higher uncertainty. Interestingly, although both statistics increase with depth, the rates of increase are different and their ratio actually decreases, causing the relative uncertainty to decrease with depth (Figure 3b). From these results, it could be concluded that, with careful uncertainty analysis of the system parameters, the mean lifetime expectancy can be used as a safety indicator in the selection of subsurface repository sites in a fractured crystalline environment.

# 4. Illustrative Example

In this section, the utility and the applicability of the concepts of lifetime expectancy and backward-in-time transport for optimal location of subsurface waste repositories are illustrated with a set of numerical examples. This paper focuses especially on the effects of fracture network geometry on lifetime expectancy, which is intended as a preliminary study for the application of the approach in the Canadian Shield environment, where interconnected fracture networks could compromise the safety of waste repositories [*Park et al.*, 2007]. Numerical simulations have been performed using a finite element groundwater flow, mass/head transport simulator [*Cornaton*, 2007]. Mean lifetime expectancy is solved in a standard way by solving the steady state backward equation (9) for both discrete fracture zone and matrix domains. It is noted that a discrete fracture dual permeability conceptualization is adapted in this study and thus a fracture (zone) represents a distinct preferential pathway in the domain and the matrix domain is less permeable fractured rock domain [*Therrien and Sudicky*, 1996]. MLE is prescribed as a homogeneous boundary condition ($E = 0$) at outlet nodes and a free mass exit condition is used at inlet nodes. Note that lifetime expectancy PDF solutions presented in the following sections are results of transient simulation of (6).

## 4.1. Model Layout and Proof-of-Concepts

Four sets of different fracture network geometries are considered in a two-dimensional, 10km wide and 1km deep, cross-sectional domain (Figure 4) and the domain is discretized into uniform 10m×10m rectangular elements. For the illustrative schematic fractured porous media, the fracture domain is assumed to be more permeable and have a higher porosity than the matrix domain. Flow and transport properties are homogeneous, with the fracture domain having a hydraulic conductivity of $10^{-3}$ m/s, porosity of 0.3, and dispersivity of 10m, and the matrix domain having an isotropic hydraulic conductivity of $10^{-6}$ m/s, porosity of 0.2, and longitudinal and transverse dispersivities of 10m and 1m respectively. Free solution diffusion coefficient is set to be $2.3\times10^{-9}$ m$^2$/s. These material properties are chosen arbitrarily here for illustrative purposes only and are not intended to be representative of a typical crystalline rock. Decay and retardation are excluded for simplification and clarity. A specified hydraulic head boundary condition is imposed at the top of the domain, to derive the steady flow from top left to top right as shown in Figure 4a.

As a first step, mean lifetime expectancy is computed throughout the whole domain for the homogeneous matrix system (Figure 5a). We would emphasize again that this result can be obtained in one step by solving the steady-state backward-in-time mean lifetime expectancy equation (9), which is equivalent to solving the forward transport equation (3), or its first moment form, a number of times using a Dirac input at each nodal upstream point. The result in Figure 5a can easily be understood from the flow field in Figure 4a: mean lifetime expectancy decreases as any point becomes closer to the top right exit boundary because one can expect a short travel time to the exit boundary if one is near the exit boundary, while travel time is greater for the bottom left corner of the domain (up to 100,000 years). With this result in mind, one might conclude that the bottom left corner would be the optimal and safest location for hosting the repository in terms of the longest average expected travel time to the biosphere.

Figure 5b shows nine probability density functions for lifetime expectancy at the observation points R11-R33 in Figure 5a. Since the probability density for lifetime expectancy accounts for the temporal probabilistic distribution for lifetime expectancy, travel time from one point within the domain to the exit boundary could vary significantly, depending on the flow field and hydro-dispersive properties of the medium. For this specific example, water molecules at the observation point, for example R22, will reach the top of the domain (biosphere) in about 30,000 years on average, but it could be 5,000 years earlier or later as shown in Figure 5b. Note that as lifetime expectancy becomes greater, it tends to exhibit greater variance as in classical transport theory.

Figure 5c shows vertical profiles of mean lifetime expectancy through the observation points R11-R13 (R1), R21-R23 (R2), and R31-R33 (R3). The results show that the expected travel time is greater as a source becomes deeper. Interestingly, the three vertical profiles in Figure 5c behave in different ways due to their different locations in the flow field. For the profile across R1, mean lifetime expectancy linearly increases with depth, while it increases quickly down to 100m depth for the profile across R2 and it increases relatively slowly down to 900m depth for R3.

## 4.2. Effects of Fracture Zone Geometry

In order to investigate the influence of fracture zone geometry on lifetime expectancy in fractured porous media, three sets of fracture network geometries are embedded in the matrix system. First we consider the case with one horizontal fracture zone at the bottom and four evenly spaced vertical fracture zones (Figure 4b) and then the cases with different numbers of vertical and horizontal fracture zones (Figure 4c and 4d).

For the case in Figure 4b, the mean lifetime expectancy in the matrix domain varies up to about 60,000 years along the recharge area and the contours tend to be refracted across fracture zones due to a higher flow rate in the discontinuities (Figure 6a). The main difference between Figures 5a and 6a lies in the mean lifetime expectancy near the bottom horizontal fracture zone: the closer to the horizontal fracture zone, the mean lifetime expectancy becomes smaller (*e.g.*, see the 5,000 year contour). Computed probability density functions of lifetime expectancy from nine observation points indicate that lifetime expectancy becomes much smaller over the domain compared to the results in Figure 5b, because fracture zones act as short cuts. Not surprisingly, the probability density for R1 has notable multiple peaks due to multiple distinct flow paths to the discharge area along the fracture zones. It is worth noting that the first peaks for R1 arrive earlier than peaks for R2 because the observation points R1 are located closer to fracture zones. The mean lifetime expectancy for R1 ranges from about 20,000 years for R13 to 45,000 years for R11, as attested to by the strong gradient in mean lifetime expectancy in the neighborhood of R1 (Figure 6a), even though the first peaks will appear between 10,000 and 20,000 years. This discrepancy is important in the context of risk or safety in hosting repositories because the weighted average time may not be an appropriate representation of risk. For example, if we would predict two equally probable risks occurring in one week or in 10 years from the present, 5 years (the arithmetic average) might not be a representative value for those two risks and it might be more prudent to worry about the earlier risk. Due to the existence of the bottom fracture zone, mean lifetime expectancy decreases near the bottom of the domain (Figure 6c). A rapid decrease in mean lifetime expectancy along R1 could be explained by downward flow in the matrix block, compared to lateral or upward directional flow in other blocks.

Figures 7 and 8 show the results for mean lifetime expectancy and the corresponding PDFs when the number of vertical or horizontal fracture zones is increased (Figures 4c and 4d). For Figures 7b and 7c, lifetime expectancy is smaller than in the previous case (Figure 6), although the general tendency in the distribution does not change significantly as the distance from the observation points to the nearest fracture zones does not change significantly. Note again that the points in R2 could represent safer repository locations than R1, regarding the first arrival times. This result could be contradictory to the suggestion of *Tóth and Sheng* [1996] that a repository could be best hosted in a recharge area. When a horizontal fracture zone is added in the middle of the domain (Figure 4d), the nine observation points become significantly closer to a fracture zone (R12, R22, and R32 are now located on the fracture zone). The mean lifetime expectancy distribution shows that the additional fracture zone could shorten the lifetime expectancy values to less than one half of their original values for most of the domain (Figures 8a and 8b). The lifetime expectancy PDFs for R12, R22, and R32 show multiple peaks and significant tailing due to matrix diffusion.

Analyses of lifetime expectancy for the illustrative examples indicate that one can expect shorter lifetime expectancy near horizontal fracture zones, and that, as a matter of fact, depth is not necessarily a secure factor in ensuring safety in fractured porous media. In addition, the comparison between the mean and PDF solutions in lifetime expectancy leads to the suggestion that multiple pathways within reactive transport systems, such as fractured porous media, will require careful analyses. This is because statistics such as temporal moments are not be sufficiently representative to capture multiple risks occurring as distinct temporal events (e.g. in the case where the lifetime expectancy PDF is multi-modal).

## 5. Discussion and Remarks

The risk associated with subsurface repositories relies upon the dose one can expect at the biosphere and the travel time to the biosphere from the repositories, combined with an acceptable prediction uncertainty. In this study, we propose a methodology for the performance assessment of radioactive waste repositories based on the analysis of the lifetime expectancy from the repository location to the biosphere. The suggested methodology can provide the theoretical foundation for selecting optimal repository locations in terms of the longest expected travel time.

Safety for a radioactive waste repository is usually affected by a threshold radioactivity (*e.g.*, expressed in Bq/l, Bq/year or Sieverts as given by EU regulations [*The Council of European Union*, 1998]). Longer travel time cannot always guarantee less risk at the biosphere. The risk associated with radioactivity releases at the biosphere, caused by potential nuclide releases from a repository site, can be quantified as described in Section 2.4. An alternative for the risk analysis is the evaluation of a risk index. Environmental risk is defined as the probability that harm will occur from the exposure to a contaminant. Risk is commonly interpreted as the fraction of the population that will experience increased health risk, which is equivalent to the increased risk to an individual in ergodic populations. Guidelines for such risk evaluation can be found, for example, in *Maxwell et al*. [1998] or *Andričevic et al*. [1994], where the risk is related to the statistics of the flux-averaged concentration. Once the lifetime expectancy PDF is computed over the domain, radioactivity associated with a release function $m^*(x,t)$ from a repository at any arbitrary location and with any arbitrary dimension can be computed by enforcing equation (10), and further converted into a health risk index.

The results for the illustrative examples showed that lifetime expectancy could be the longest near the surface in the recharge area, where flow direction is downward to the subsurface. This might lead to the idea that a repository could be hosted at a recharge area, as proposed by *Tóth and Sheng* [1996] for a Canadian Shield environment. However, near surface the flow field is highly uncertain because it is subject to rapid fluctuations in external stresses, such as climate. Thus, any location near surface is possibly subject to the highest uncertainty and should thus be avoided to host a repository, contrary to the suggestions provided by *Tóth and Sheng* [1996]. Moreover, the theoretical results carried out in the example problems clearly show how fracture zones can represent conductive pathways that can drastically reduce the travel times from a repository to the biosphere. Thus, we would stress that the uncertainty in the reservoir's structure and parameter distributions leads to uncertainty in predicting lifetime expectancy and these aspects should be considered in real application problems.

This study is a preliminary phase for future enhanced simulations and applications. It allows the analysis of uncertain variables and provides a framework to isolate safe repository locations, which can be further tested and constrained by incorporating specific physical and chemical processes. Further developments will include: effect of shield brine distributions on radionuclide transport; the effect of the variations of climatic and hydraulic conditions (e.g. effects of potential future glaciation events); coupling with heat transport (thermo-haline processes); multi-component reactive transport of radionuclides. In a subsequent article [*Park et al.*, 2007] we discuss the risk prediction uncertainty for the application of the proposed methodology in a typical fractured crystalline Canadian Shield environment.

**Acknowledgments**

This research was supported by Ontario Power Generation (OPG), the Natural Sciences and Engineering Research Council (NSERC) of Canada, and by the Swiss Research National Fund (Grant no. 2100-064927) to F. J. Cornaton. Additional funding was provided by grant 3-2-3 from Sustainable Water Resources Research Center of the 21$^{st}$ Century Frontier Research Program of Korea.

Goode, D. J. (1996), Direct simulation of groundwater age, *Water Resour. Res.*, 32, 289–296.

Jury, W. A. (1982), Simulation of solute transport using a transfer function model, *Water Resour. Res.*, 18(2), 363–368.

Jury, W. A., and K. Roth (1990), *Transfer Functions and Solute Transport through Soil: Theory and Applications*, Birkhauser Publ., Basel, Germany.

Kolmogorov, A. N. (1931), Űber die analytischen methoden in der wahrscheilichkeitsrechnung, *Math. Anal.*, 104, 415–458.

Van Kooten, J. J. A. (1995), An asymptotic method for predicting the contamination of a pumping well, *Adv. Water Res.*, 18(5), 295–313.

LaBolle, E. M., J. Quastel , and G. E. Fogg (1998), Diffusion theory for transport in porous media: Transition-probability densities of diffusion processes corresponding to advection--dispersion equations, *Water Resour. Res.*, 34(7), 1685–1693.

LaBolle, E. M., J. Quastel, G. E. Fogg, and J. Gravner (2000), Diffusion processes in composite porous media and their numerical integration by random-walks: Generalized stochastic differential equations with discontinuous coefficients, *Water Resour. Res.*, 36(3), 651–662.

Langevin, P. (1908), Sur la théorie du mouvement brownien, *Comptes Rendus de l'Académie des Sciences*, 146, 530–533.

Maxwell, R. M., S. D. Pelmulder, A. F. B. Tompson, and W. E. Kastenberg (1998), On the development of a new methodology for groundwater-driven health risk assessment, *Water Resour. Res.*, 34(4), 833–847.
28

**Figure Captions**

**Figure 1.** Schematics for a set of parallel vertical fractures and matrix block system, to derive analytical solutions for temporal moments for lifetime expectancy.

**Figure 2.** Mean lifetime expectancy solutions for a set of parallel fractures and matrix block system in Figure 1.

**Figure 3.** (a) Mean ($E_f$) and standard deviation ($\sigma_f$) in lifetime expectancy and (b) the coefficient of variation ($\sigma_f / E_f$) with depth for fracture domain.

**Figure 4.** Schematics of the vertical cross-sections used to model mean and probability density for lifetime expectancy: (a) hydraulic head distribution in a homogeneous matrix system under given flow boundary conditions; (b)-(d) variation cases with different fracture networks (thick lines).

**Figure 5.** Lifetime expectancy solutions for the case in Figure 4a. (a) Mean lifetime expectancy distribution in years, (b) lifetime expectancy probability density at nine observation points in (a), and (c) vertical logs of mean lifetime expectancy passing through the observation points.

**Figure 6.** Lifetime expectancy solutions for the case in Figure 4b. (a) Mean lifetime expectancy distribution in years, (b) lifetime expectancy probability density at nine observation points in (a), and (c) vertical logs of mean lifetime expectancy passing through the observation points.

**Figure 7.** Lifetime expectancy solutions for different fracture network geometries. (a) and (c) show lifetime expectancy probability density at nine observation points for Figures 4c and 4d, respectively. (b) and (d) are vertical logs of mean lifetime expectancy passing through the observation points for Figures 4c and 4d.

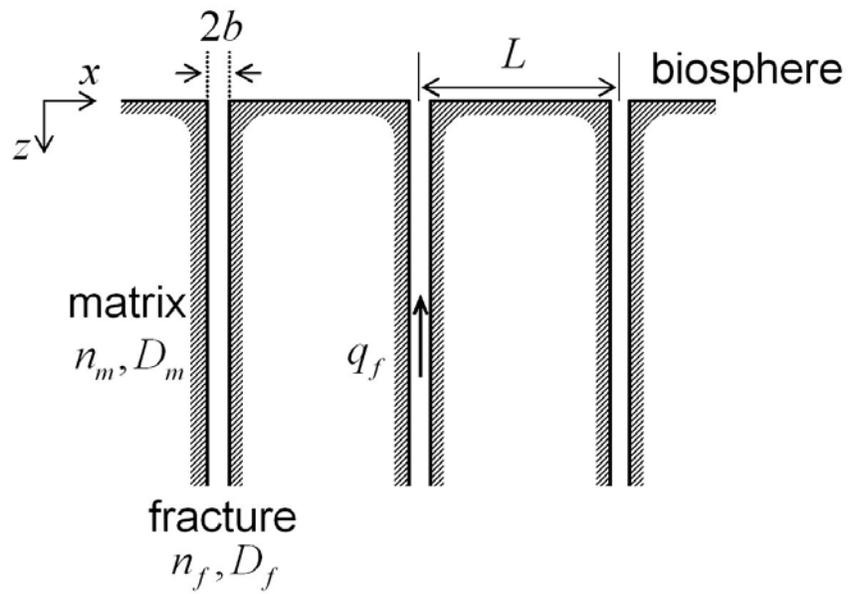

**Figure 1.** Schematics for a set of parallel vertical fractures and matrix block system, to derive analytical solutions for temporal moments for lifetime expectancy.

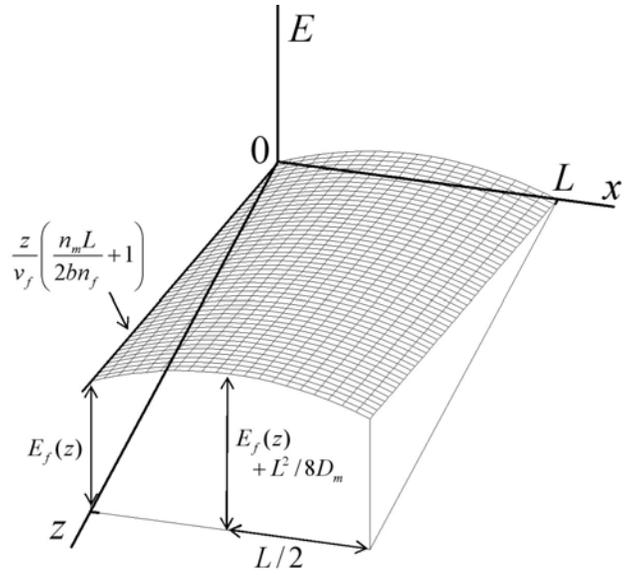

**Figure 2.** Mean lifetime expectancy solutions for a set of parallel fractures and matrix block system in Figure 1.

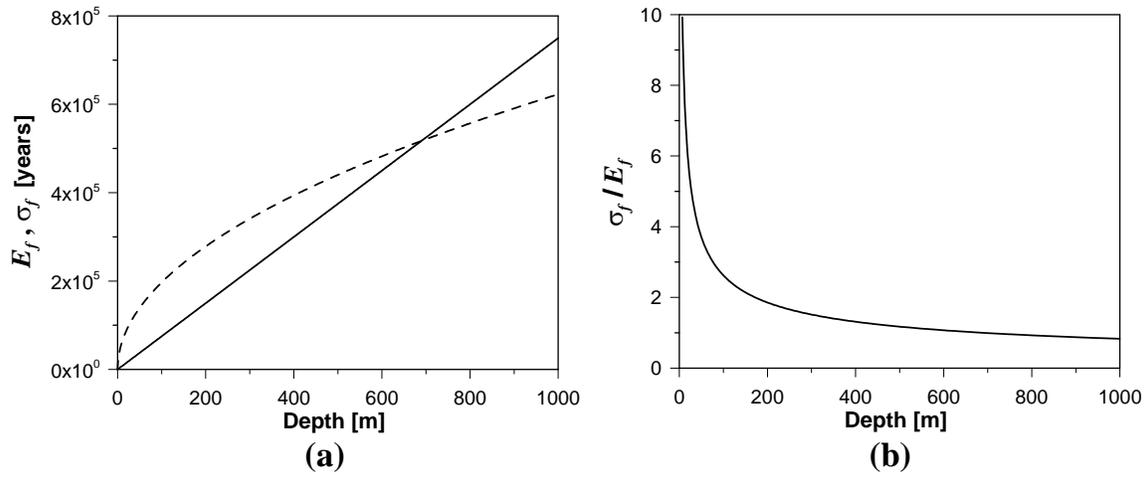

**Figure 3.** (a) Mean ($E_f$: solid line) and standard deviation ($\sigma_f$: dashed line) in lifetime expectancy and (b) the coefficient of variation ($\sigma_f / E_f$) with depth for fracture domain.

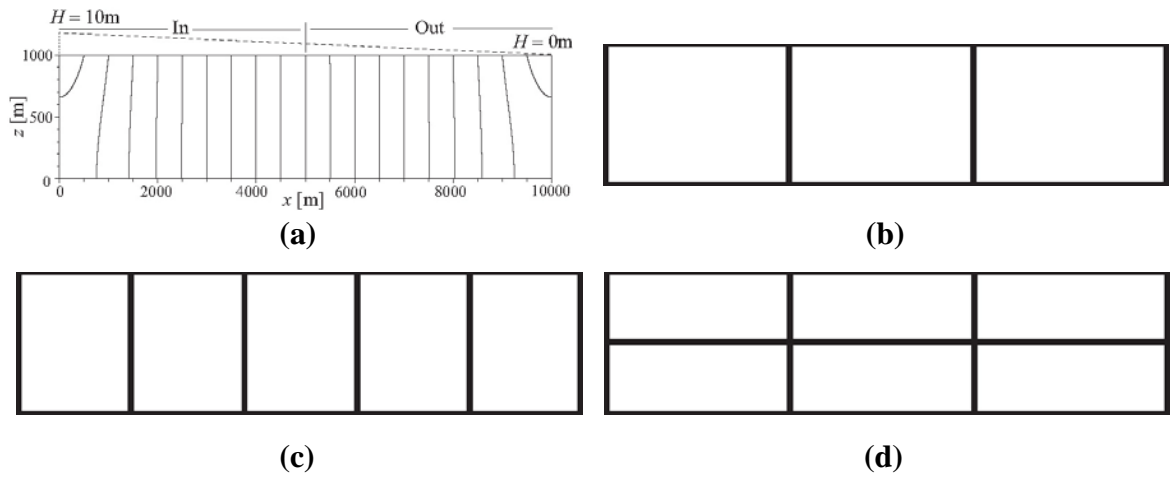

**Figure 4.** Schematics of the vertical cross-sections used to model mean and probability density for lifetime expectancy: (a) hydraulic head distribution in a homogeneous matrix system under given flow boundary conditions; (b)-(d) variation cases with different fracture networks (thick lines).

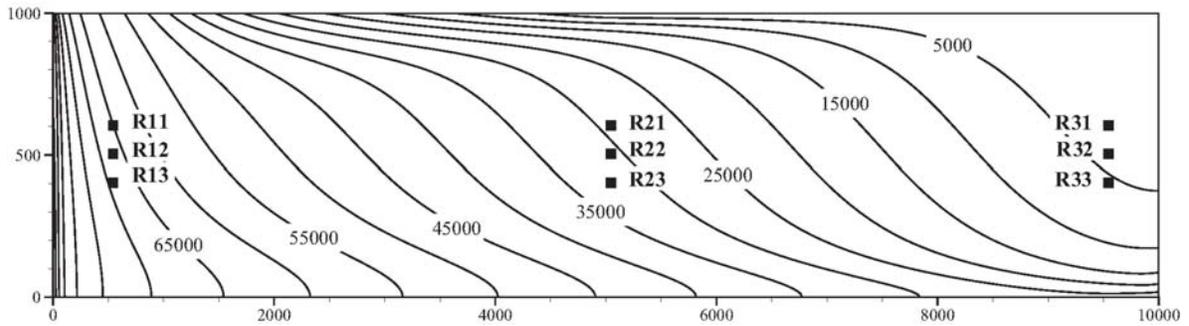
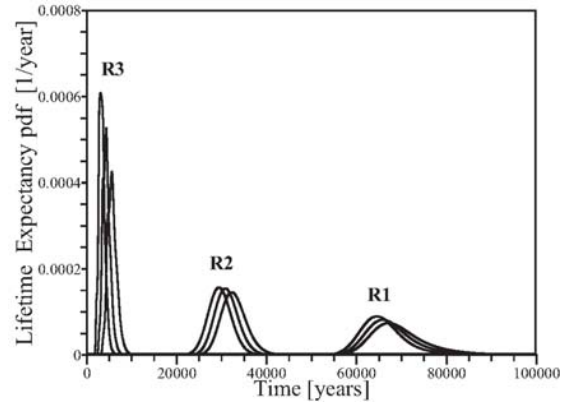
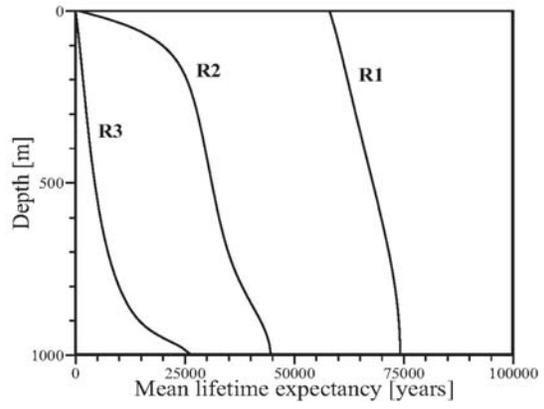

**Figure 5.** Lifetime expectancy solutions for the case in Figure 4a. (a) Mean lifetime expectancy distribution in years, (b) lifetime expectancy probability density at nine observation points in (a), and (c) vertical logs of mean lifetime expectancy passing through the observation points.

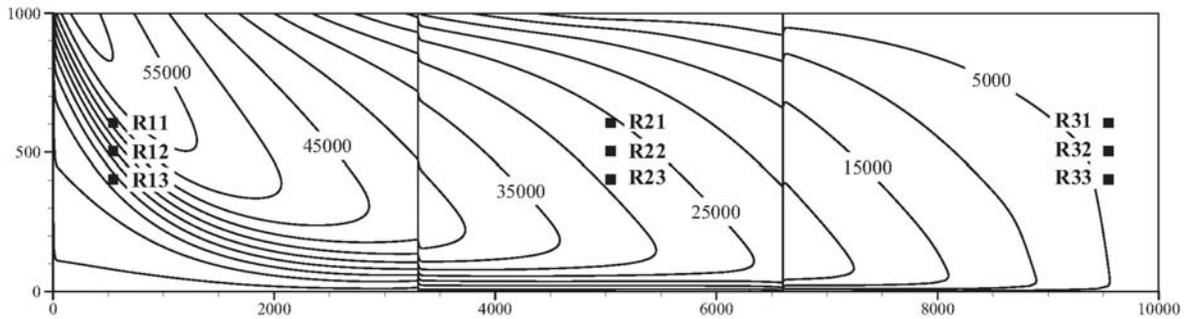

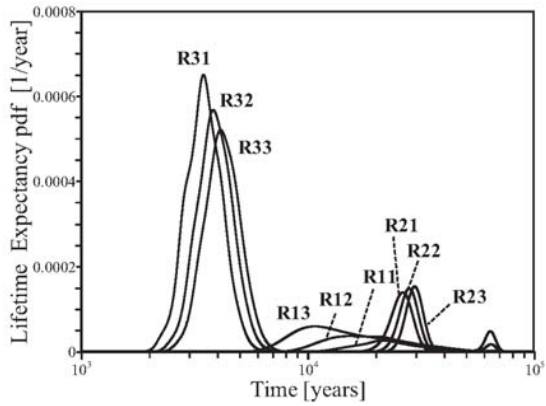
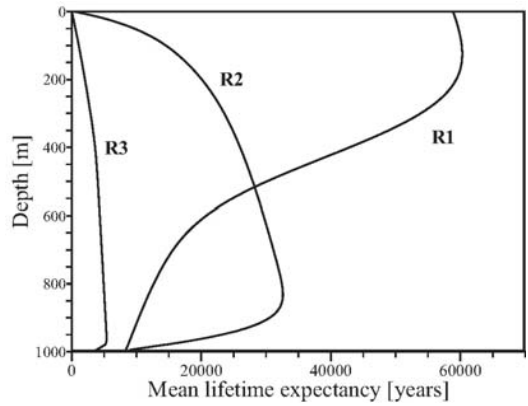

**Figure 6.** Lifetime expectancy solutions for the case in Figure 4b. (a) Mean lifetime expectancy distribution in years, (b) lifetime expectancy probability density at nine observation points in (a), and (c) vertical logs of mean lifetime expectancy passing through the observation points.

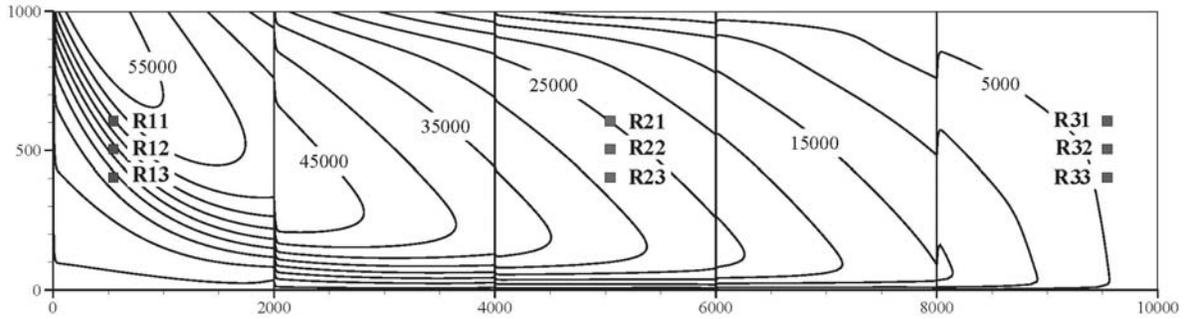

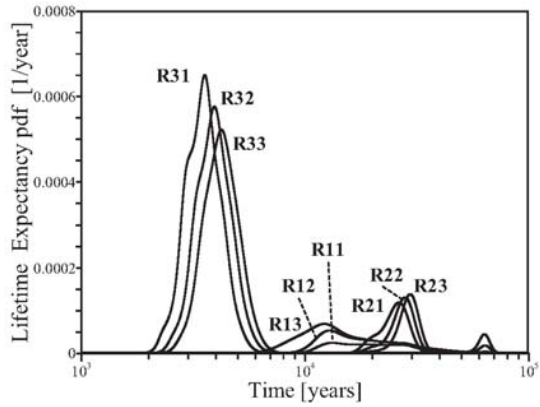
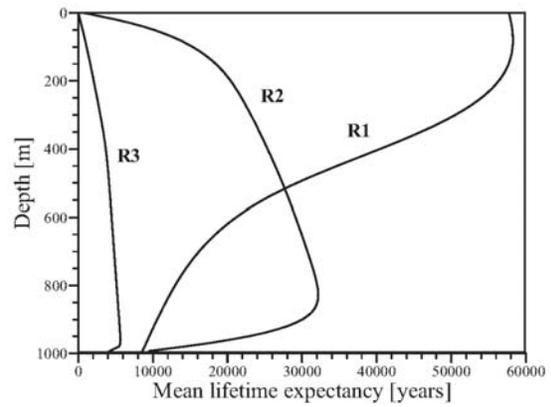

**Figure 7.** Lifetime expectancy solutions for the case in Figure 4c. (a) Mean lifetime expectancy distribution in years, (b) lifetime expectancy probability density at nine observation points in (a), and (c) vertical logs of mean lifetime expectancy passing through the observation points.

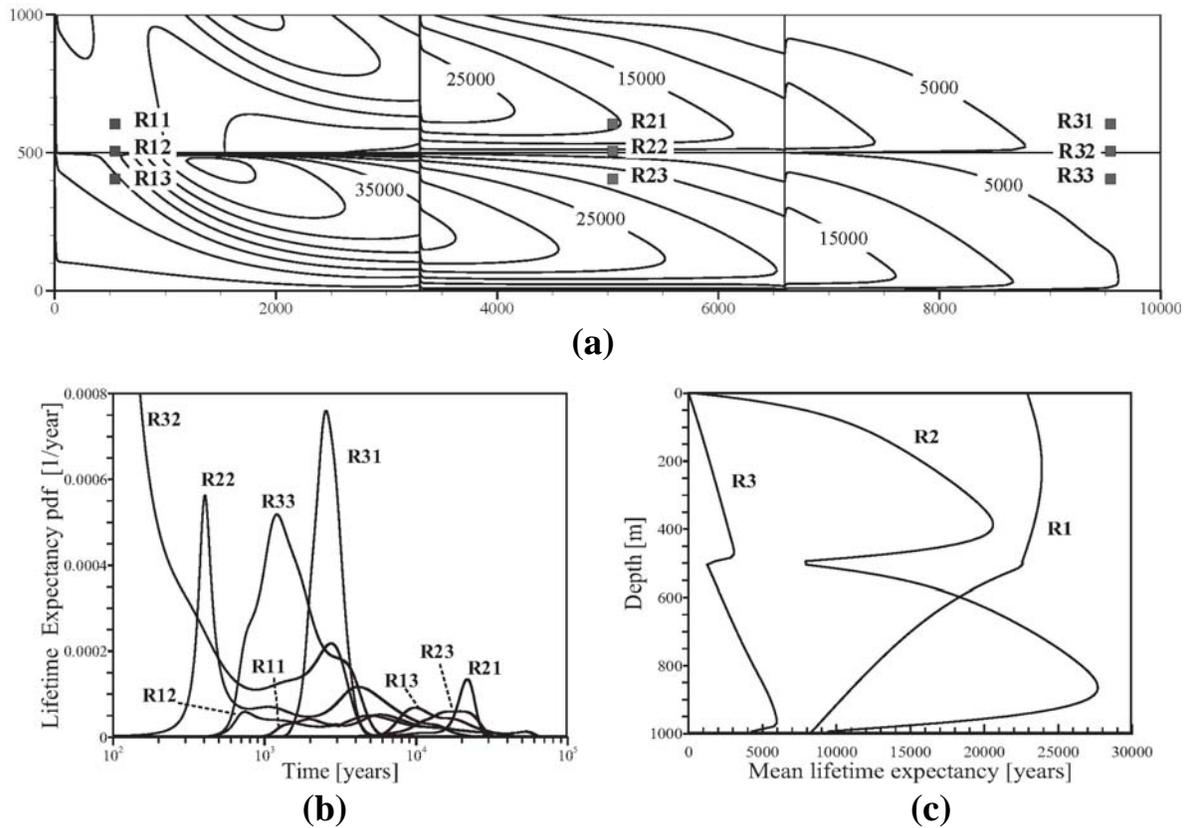

**Figure 8.** Lifetime expectancy solutions for the case in Figure 4d. (a) Mean lifetime expectancy distribution in years, (b) lifetime expectancy probability density at nine observation points in (a), and (c) vertical logs of mean lifetime expectancy passing through the observation points.